\documentclass[final]{svjour3}
\usepackage{graphicx}
\usepackage{rotating}
\usepackage{amssymb}
\usepackage{mathptmx}
\usepackage[numbers]{natbib}
\makeatletter
\journalname{Journal of Low Temperature Physics}
\bibpunct{}{}{,}{s}{}{,}

\begin{document}

\newcommand{\hdblarrow}{H\makebox[0.9ex][l]{$\downdownarrows$}-}
\title{Full-Array Noise Performance of Deployment-Grade SuperSpec
  mm-wave On-Chip Spectrometers}
\titlerunning{Noise Performance of SuperSpec On-Chip Spectrometers}

\author{K.~S.~Karkare$^1$ \and P.~S.~Barry$^{2,1}$ \and
  C.~M.~Bradford$^{3,4}$ \and S.~Chapman$^5$ \and S.~Doyle$^6$ \and
  J.~Glenn$^7$ \and S.~Gordon$^8$ \and S.~Hailey-Dunsheath$^4$ \and
  R.~M.~J.~Janssen$^{3,4}$ \and A.~Kov\'{a}cs$^9$ \and H.~G.~LeDuc$^3$
  \and P.~Mauskopf$^8$ \and R.~McGeehan$^1$ \and J.~Redford$^4$ \and
  E.~Shirokoff$^1$ \and C.~Tucker$^6$ \and J.~Wheeler$^7$ \and
  J.~Zmuidzinas$^{4}$}
\authorrunning{K.~S.~Karkare et al.}

\institute{\email{kkarkare@kicp.uchicago.edu}\\
  1: Kavli Institute for Cosmological Physics, University of Chicago, Chicago, IL 60637, USA. \\
  2: High-Energy Physics Division, Argonne National Laboratory, Argonne, IL 60439, USA\\
  3: Jet Propulsion Laboratory, Pasadena, CA 91109, USA\\
  4: California Institute of Technology, Pasadena, CA 91125, USA\\
  5: Department of Physics and Atmospheric Science, Dalhousie University, Halifax NS B3H 1A6, Canada\\
  6: School of Physics \& Astronomy, Cardiff University, Cardiff CF24 3AA, UK\\
  7: Center for Astrophysics and Space Astronomy, University of Colorado Boulder, Boulder, CO 80309, USA\\
  8: School of Earth and Space Exploration, Arizona State University, Tempe, AZ 85287, USA\\
9: Harvard-Smithsonian Center for Astrophysics, Cambridge, MA 02138}

\maketitle

\begin{abstract}
SuperSpec is an on-chip filter-bank spectrometer designed for wideband
moderate-resolution spectroscopy at millimeter wavelengths, employing
TiN kinetic inductance detectors.  SuperSpec technology will enable
large-format spectroscopic integral field units suitable for
high-redshift line intensity mapping and multi-object spectrographs.
In previous results we have demonstrated noise performance in
individual detectors suitable for photon noise limited ground-based
observations at excellent mm-wave sites.  In these proceedings we
present the noise performance of a full $R\sim 275$ spectrometer
measured using deployment-ready RF hardware and software.  We report
typical noise equivalent powers through the full device of $\sim 3
\times 10^{-16}$ W Hz$^{-1/2}$ at expected sky loadings, which are
photon noise dominated.  Based on these results, we plan to deploy a
six-spectrometer demonstration instrument to the Large Millimeter
Telescope in early 2020.


\end{abstract}

%When preparing your manuscript, please follow the instructions in this
%template. This template serves as a universal template for Special
%Issue Articles to be taken into consideration for publication in the
%Journal of Low Temperature Physics.\\ 
%Insert an empty line after the section title. Indent at the start of
%each paragraph after the first paragraph of the section, which is not
%indented. For special issue articles the page length is 6
%pages. Invited contributions may be 12 pages long. In order to
%estimate your article length, please prepare your manuscript in Times
%or Times New Roman, font size 11, justify the body text, and make sure
%the page format is set to A4. The template margins are: Top: 2.07”,
%Left: 1.77”, Bottom: 2.17”, Right: 1.77” inches.\\ 
%Subsections can be used. 

\section{Introduction}

Spectroscopy at millimeter wavelengths is a unique probe of the early
Universe.  High-redshift star-forming galaxies emit most of their
energy in the infrared due to dust absorption of optical/UV radiation
from young stars\cite{lagache05}.  In addition to dust continuum, this
radiation also excites far-IR line emission such as the CO $\mathrm{J}
\rightarrow \mathrm{J}-1$ rotational ladder and the [CII] ionized
carbon fine structure transition.  Detecting these lines---which for
high-$z$ galaxies are redshifted into the mm atmospheric window and
observable from the ground---enables precise determination of a
galaxy's redshift and unique understanding of its astrophysical
properties.

While detections have been made with ALMA\cite{bradac17}, its modest
bandwidth and narrow field of view preclude it from efficiently
surveying large numbers of galaxies.  A large-scale instrument with
moderate mm-wave spectral resolution and hundreds of detectors would
be an invaluable complement to ALMA's individual, deep spectra.
Science goals such as understanding the global star formation rate and
molecular gas content as a function of cosmic time would be enabled
with a steered multi-object spectrograph.  Since galaxies also trace
the underlying dark matter, their spatial distribution probes
cosmology and would be accessible with an integral field spectrometer
performing line intensity mapping\cite{kovetz17} of CO and [CII].
Intensity mapping experiments with filled focal planes could constrain
the expansion history\cite{karkare18} or primordial non-Gaussianity to
higher precision than galaxy surveys by extending the available
cosmological volume from $z \sim 3$ up to the Epoch of Reionization
($z \sim 9$).

SuperSpec is designed to enable the next generation of large-format
mm-wave spectroscopic surveys.  In our device a mm-wave filter bank
spectrometer patterned on a few-square-cm silicon chip provides the
favorable attributes of a diffraction grating spectrometer---wide
instantaneous bandwidth and background-limited sensitivity---but in a
package which is orders of magnitude smaller.  The design can be
scaled up to hundreds or thousands of spectrometers on modestly-sized
instruments, potentially enabling an increase in sensitivity similar
to that experienced by recent cosmic microwave background experiments.
The natural multiplexing capability of kinetic inductance detectors
(KIDs) enables the high detector counts necessary for these
instruments.  Several other projects are pursuing similar approaches,
including the DESHIMA on-chip filter bank\cite{endo19}, the Micro-Spec
on-chip grating spectrometer\cite{cataldo18}, and the WSPEC waveguide
filter bank \cite{bryan16}.

In previous publications we have described the spectrometer design,
filter performance, and noise characteristics for individual
spectrometer channels suitable for excellent mm-wave observing
sites\cite{shirokoff12, haileydunsheath16, wheeler16}.  We plan to
deploy a demonstration instrument to the 50-m Large Millimeter
Telescope\cite{schloerb04} (LMT) in early 2020.  This instrument will
feature six individual SuperSpec devices, configured as three
dual-polarization beams on the sky.  In these proceedings, we present
the noise performance of a full SuperSpec device in deployment
configuration, i.e., with the readout RF hardware and software that we
plan to use at the LMT, and with optical loading similar to on-sky
values.  The device tested is a 50-channel spectrometer fully covering
255--278\,GHz.  In Section~\ref{sec:setup} we describe the device,
optical setup, and readout configuration.  Measurement results
including full-device noise equivalent power (NEP) are presented in
Section~\ref{sec:results}, and we conclude in
Section~\ref{sec:discussion}.

\section{Experimental Setup}
\label{sec:setup}

In preparation for deployment to the LMT, we characterized a
deployment-grade SuperSpec device in several configurations, including
one designed to match astronomical observations.  Here we describe the
device and experimental setups.

\subsection{SuperSpec Device and Optical Configurations}
We tested a 3rd-generation SuperSpec device: a 50-channel spectrometer
with resolving power $R\sim 275$, spanning 255--278\,GHz and
illustrated in Fig.~1\citep{haileydunsheath16, wheeler16}.  Radiation
incident on the device is first focused by an anti-reflection coated
hyperhemispherical silicon lenslet.  It is then coupled to microstrip
using a dual bow-tie slot antenna, which achieves the requisite wide
bandwidth.  Traveling down the feedline, the mm-wave radiation is
proximity coupled to a series of tuned $\lambda/2$ filters, decreasing
in frequency.  The currents in these filters couple to the inductor of
a TiN KID.  Finally, the KID resonator circuit (with microwave
frequencies from 80--160 MHz) is coupled to the coplanar
waveguide readout line using an interdigitated capacitor.  The tested
device has been re-etched to separate 6 pairs of resonators that were
previously collided\cite{mcgeehan18} (separated by less than 5
linewidths).  The typical $T_c$ is $\sim 0.93$ K.  The device was held
at a base temperature of 230 mK, achievable with a standard
$^4$He-$^3$He-$^3$He sorption refrigerator.

\begin{figure}
\begin{center}
\includegraphics[width=0.9\linewidth, keepaspectratio]{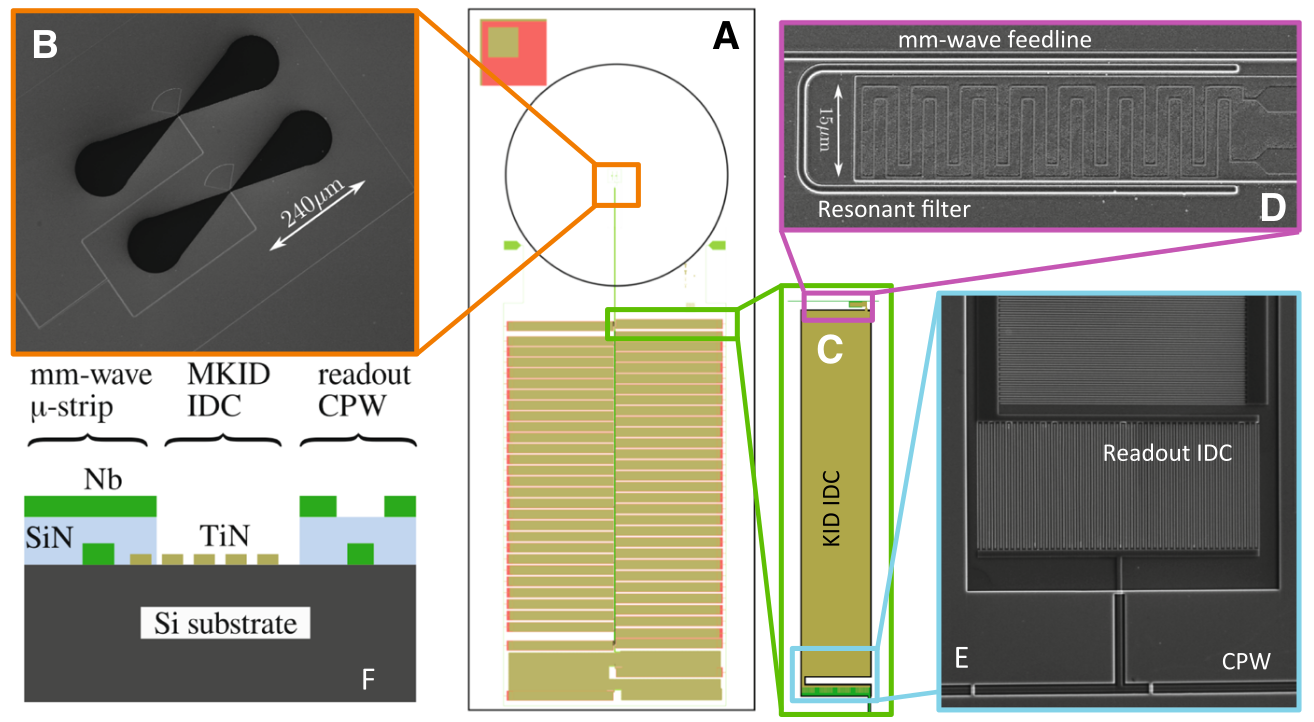}
\caption{50-channel SuperSpec device\cite{wheeler18}.  \textbf{a}:
  Mask overview with lens footprint/antenna at the top, and feedline
  running vertically past the filters.  \textbf{b}: Slot antenna.
  \textbf{c}: A mm-wave filter and KID.  \textbf{d}: The $\lambda/2$
  mm-wave resonator and KID inductor.  \textbf{e}: The lower portion
  of the KID interdigitated capacitor (IDC), coupling IDC, and readout
  line.  \textbf{f}: Cross-section of device layers.  (Color figure
  online.)}
\end{center}
\label{fig:ss_image}
\end{figure}

A swept coherent source was used to measure the millimeter-wave filter
bank at high spectral resolution.  The 220--330\, GHz source,
consisting of a commercial amplifier/multiplier chain, was coupled to
a feed horn and radiated directly into the optical cryostat
\cite{haileydunsheath14}.  For noise measurements, a low-loading
environment was used to simulate observing at the LMT.  Using
historical precipitable water vapor values measured
on-site\cite{zeballos16} we calculated the expected loading incident
on the device lens (using the zenith temperature predicted by the
\textit{am} atmospheric model\cite{am}, the measured mirror surface
roughness, and other estimates of illumination efficiency).  In
typical conditions we will likely see $\sim 0.7$ pW of loading
incident on the lenslet, depending on the detector bandwidth.  We
therefore used a cryogenic blackbody load, which filled the beam
emerging from the lenslet.  The load was coupled to the 4 K stage of
the pulse tube cooler, and had a temperature variable up to 60 K
($\sim 1.5$ pW).

\subsection{RF Chain and Readout}
The device is read out with a ROACH2-based system using firmware that
was originally developed for the BLAST-TNG experiment\cite{gordon16}.
A comb of resonant frequencies is used to synthesize a waveform which
is output from the ROACH2 FPGA to a MUSIC ADC/DAC board.  The baseband
frequencies range from $\pm 256$ MHz.  The DAC output is then mixed
with a local oscillator (LO) to the KID frequencies.  We drive the
KIDs at the optimal power (typically 0.5--1 dB below bifurcation,
which varies by $\sim 2$ dB across the filter bank) by adjusting the
individual tone powers in the DAC waveform and using a variable
attenuator at the cryostat input.  After interacting with the
detectors, a low-noise amplifier at 4K amplifies the signal by 40 dB.
Upon exiting the cryostat, another variable attenuator is used to keep
the output signal within the full-scale range of the ADC.  The signal
is then demodulated using a copy of the LO signal.  The now-baseband
frequencies are finally digitized by the ADC.  The firmware filters,
Fourier transforms, digitally down-converts, downsamples, and
packetizes the $I$ and $Q$ timestreams.  The downsampled data can be
saved at a maximum rate of 488 Hz.  For our scan strategy, which will
use a chopping mirror operating at a few Hz, we may downsample
further.

To control the ROACH2 boards and RF hardware, we have developed a
Python-based software suite called \textit{pcp} (Python control
program).  \textit{pcp} contains the functionality to perform
frequency sweeps, adjust and fine-tune probe tone frequencies,
optimize readout power levels, start and stop data streaming, and
live-stream detector timestreams along with auxiliary data (such as
telescope pointing or chopping mirror position).  All six ROACH2
boards are controllable simultaneously, and depending on the range of
readout frequencies may use a single or multiple LOs.  The noise data
in Section~\ref{sec:results} were taken with \textit{pcp}.

\section{Measurement Results}
\label{sec:results}

\subsection{Filter Bank, Resonant Frequencies, and Responsivity}

\begin{figure}
\begin{center}
\includegraphics[width=1.1\linewidth, keepaspectratio]{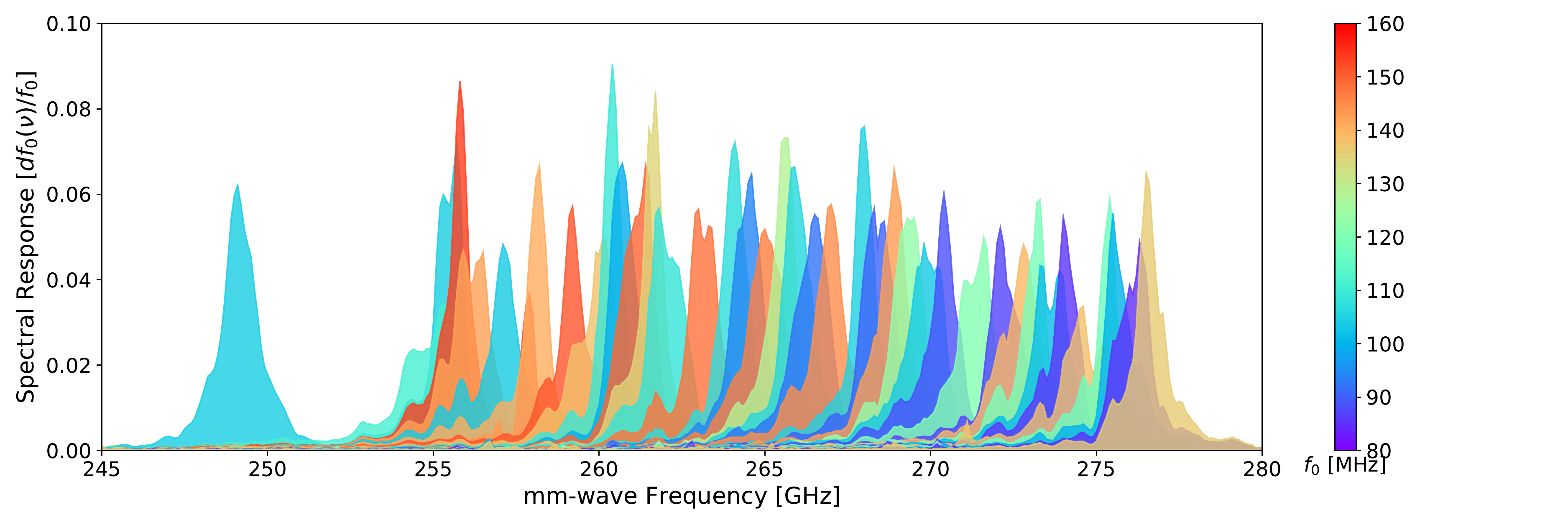}
\caption{Spectral profiles of a 50-channel SuperSpec device measured
  with an amplified coherent source.  Each profile is
  integral-normalized and has resolving power $R \sim 275$.  The
  profiles are colored according to microwave readout frequency $f_0$
  (colorbar).  Resonators that have similar mm-wave response are
  widely spaced out in readout frequency.  (Color figure online.)}
\end{center}
\label{fig:spectra}
\end{figure}

Spectral profiles of the 50-channel device are shown in Fig.~2,
measured by sweeping the coherent source across the bandwidth of the
filter bank.  The typical FWHM of these Lorentzian profiles is $\sim
1$ GHz, as expected for the $R\sim 275$ filter bank.  Profiles are
colored by microwave readout frequency $f_0$, which range from 80--160
MHz.  Note that by design, resonators with similar mm-wave response
are widely spaced out in readout frequency\cite{shirokoff12}.

Responsivity to an optical load was determined by measuring the shift
in resonant frequency as the cryogenic blackbody load temperature was
varied.  The power incident on the device was calculated as $P = \eta
\lambda^2 B(\nu,T)\Delta \nu$. $\eta = 0.65$ accounts for optical
filters between the device and the blackbody; their combined
transmission spectrum is flat to within several percent across the
255--278\,GHz band.  The bandwidth is determined directly from the
filter profiles $S(\nu)$, shown in Fig.~2, as $\Delta \nu = \left(
\int S(\nu)\ d\nu \right)^2 / \int S^2(\nu) \ d\nu$.  The typical
bandwidth is $\Delta \nu \sim 3$ GHz.  Fig.~3 shows the fractional
frequency shift $df_0 / f_0$ vs. load for each operational KID,
colored by mm-wave frequency.  Dark KIDs---i.e., those not coupled
through the filter bank---exhibited nonzero response to the blackbody
($\sim 40\%$ of the typical filter bank KID response), which was
subtracted from the optical devices assuming constant responsivity
across the tested power range.  The mean responsivity is $R = (1.8 \pm
0.6) \times 10^8$ W$^{-1}$, where the error indicates the typical
KID-to-KID variation and not measurement uncertainty.  While in a
previous publication we reported a \textit{detector-only} responsivity
referenced to power absorbed at the KID\cite{wheeler18}, this
\textit{full-device} responsivity is referenced to power incident on
the lens and includes losses in the lens, antenna, and filter bank.

\begin{figure}
\begin{center}
\includegraphics[width=1.0\linewidth, keepaspectratio]{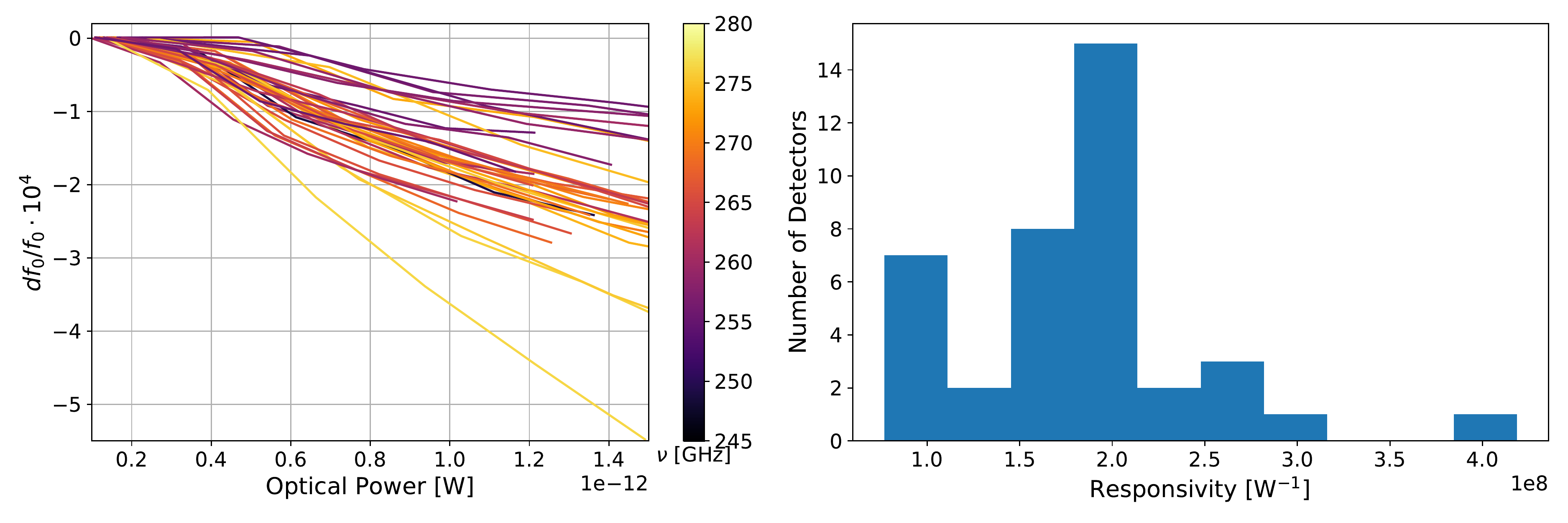}
\caption{Left: Fractional frequency shift ($df_0/f_0$) vs incident
  optical load ($P$) as the cryogenic blackbody temperature is varied.
  Curves are colored according to the peak mm-wave frequency
  (colorbar).  Power is calculated as incident on the lenslet; the
  variation in power is primarily due to differences in detector
  bandwidths.  Right: Histogram of responsivities derived from the
  slopes of the left-hand plot.  (Color figure online.)}
\end{center}
\label{fig:ffs_vs_P}
\end{figure}

\subsection{Power Spectral Densities and Noise Equivalent Power}

Noise performance in conditions similar to those at the LMT was
measured using deployment-grade RF hardware and \textit{pcp}.  At two
cryogenic blackbody temperatures corresponding to realistic sky
loadings (0.7 and 1.2 pW for the average KID), we tuned tone
frequencies and readout powers using \textit{pcp}'s automated routines
and then recorded detector timestreams at the full 488 Hz data rate.
Previously we have shown that added noise from the readout system is
negligible when the detectors are on-resonance\cite{mcgeehan18}.
There was non-negligible response to the pulse tube cooler at 1.4 Hz,
which was notch-filtered from the timestreams.

Fractional frequency power spectral densities (PSDs) $S_{xx}$ for all
channels are shown in Fig.~4 at the two optical loadings.  There is
common-mode noise across all detectors on long timescales ($1/f$ knee
at $\sim 0.7 $ Hz), which is dominated by temperature drifts from the
PID loop holding the device at 230\,mK.  Performing a principal
component analysis to remove correlated noise reduces this to $\sim
0.3$ Hz and softens the slope (Fig.~4 green curves); the white noise
level changes by only $\sim 3\%$.  At the LMT we plan to observe by
chopping on and off source with a mirror at $>1$ Hz, well into the
white noise regime of the KIDs.

\begin{figure}
\begin{center}
\includegraphics[width=1.0\linewidth, keepaspectratio]{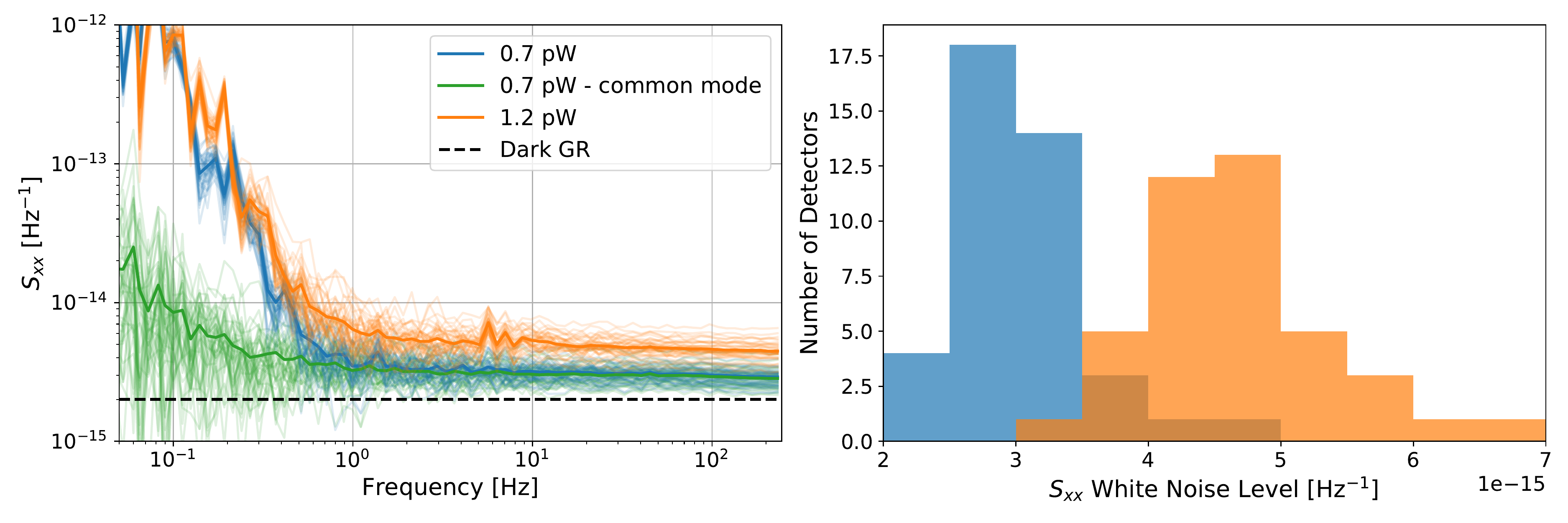}
\caption{Left: Power spectral densities $S_{xx}$ for all spectral
  channels in fractional frequency units, for an optical loading of
  $\sim 0.7$ pW (blue) and $\sim 1.2$ pW (orange).  The green curves
  show the 0.7 pW PSDs after common-mode noise removal.  The average
  noise level in a dark setting, dominated by GR noise, is also
  indicated\cite{mcgeehan18}.  Pulse tube pickup has been filtered
  out.  Amplifier noise, measured with several blind tones
  (off-resonance), is subdominant.  Right: Histogram of the white
  noise levels derived by averaging PSDs in the left-hand plot for all
  frequencies above 1 Hz.  (Color figure online.)}
\end{center}
\label{fig:sxx}
\end{figure}

In previous testing of the same device we measured the noise in a dark
environment to be $S_{xx} \sim 2 \times 10^{-15}$ Hz$^{-1}$ for the
average channel, which is consistent with being dominated by
generation-recombination (GR) noise\cite{mcgeehan18}.  Here, we see
that the addition of an optical load leads to an increment over this
value (shown with a dashed black line in Fig.~4) that is
proportional to the load.  Converting to noise equivalent power (NEP)
with $\mbox{NEP} = \sqrt{S_{xx}}/R$, where $S_{xx}$ is calculated as
the average of the white-noise regime of the PSDs ($f > 1$ Hz), we
obtain the distributions in Fig~5.  The array means are $(3.4 \pm
1.3)\times 10^{-16}$ W\,Hz$^{-1/2}$ and $(4.2 \pm 1.6) \times
10^{-16}$ W\,Hz$^{-1/2}$ for 0.7 and 1.2 pW, respectively.

Modeling the measured NEP as the sum of the dark GR noise and photon
noise, i.e., $\mathrm{NEP}_{\mathrm{meas}} =
\sqrt{\mathrm{NEP}^2_{\mathrm{GR}} + \mathrm{NEP}^2_{\mathrm{ph}}}$,
we obtain averages of $\mathrm{NEP}_{\mathrm{ph}} = 2.6 \times
10^{-16}$ W\,Hz$^{-1/2}$ for 0.7 pW and $3.6 \times 10^{-16}$
W\,Hz$^{-1/2}$ for 1.2 pW.  Comparing to the theoretical expectation
for photon noise at these loadings and accounting for a
power-dependent recombination term in the photon NEP, the optical
efficiencies are 9\% and 10\%---consistent with our simulations for a
single detector in this filter bank (note that at any one frequency,
oversampling of spectral profiles ensures that the \textit{total}
efficiency is much higher).  However, given the number of assumptions
(e.g., that of a beam-filling load) there is substantial uncertainty
on these estimates.  In this loading range the photon contribution
overtakes the GR component; therefore, at higher loadings the device
is photon noise dominated.  We again note that these
\textit{full-device} responsivities, NEPs, and efficiencies are
referenced to power incident on the device lens (but not outside the
cryostat).  They include losses from the lens, antenna, and filter
bank, and are not directly comparable to the \textit{detector-only}
values reported in other publications\cite{wheeler18} which are
referenced to power absorbed at the KID.

\begin{figure}
\begin{center}
\includegraphics[width=0.7\linewidth, keepaspectratio]{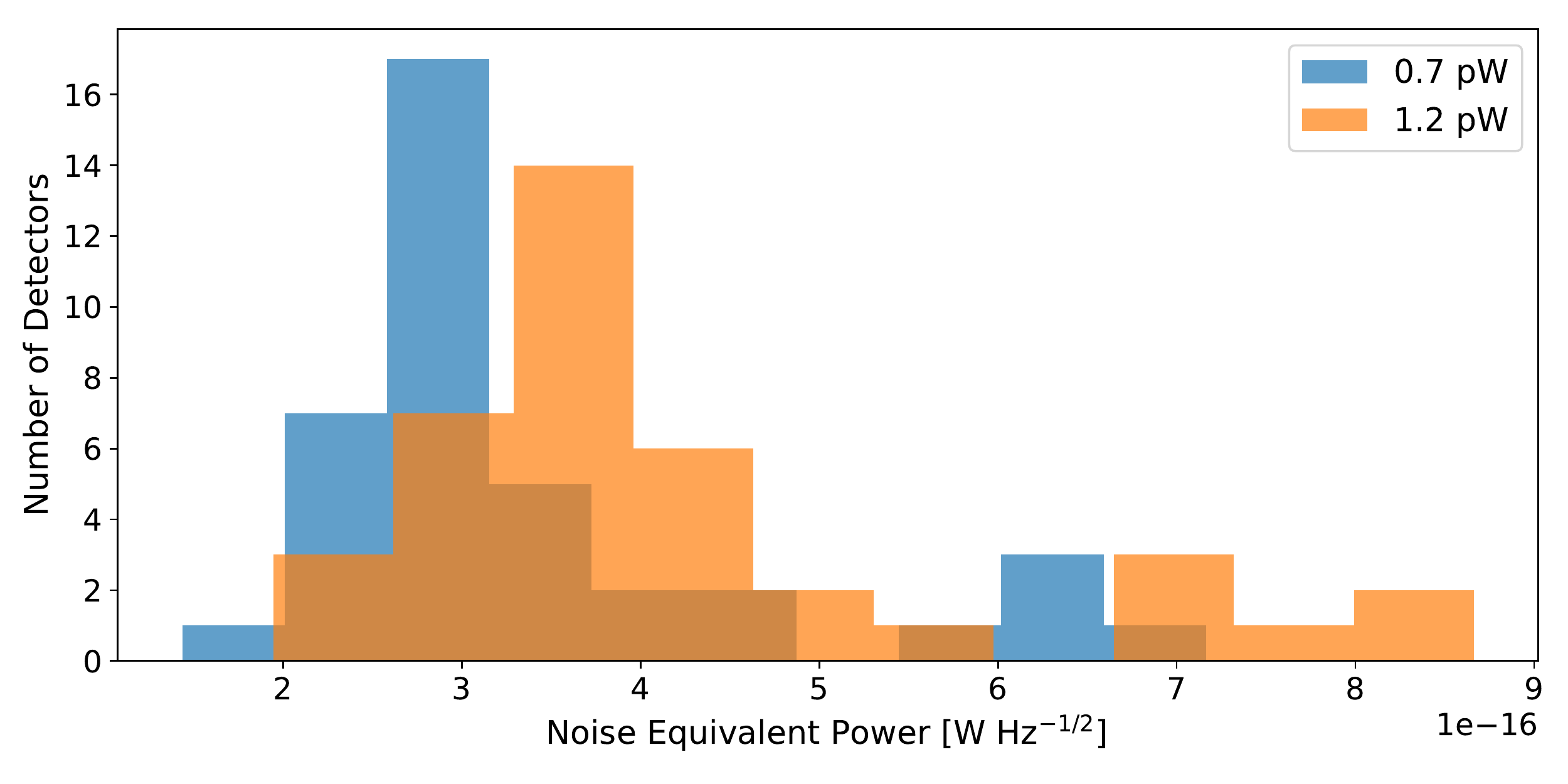}
\caption{Histograms of full-device NEPs for all spectral channels,
  calculated from the white-noise portion of the PSDs and
  responsivity, for optical loadings of $\sim 0.7$ pW (blue) and $\sim
  1.2$ pW (orange).  These NEPs include losses from the lens, antenna,
  and filter bank.  (Color figure online.)}
\end{center}
\label{nep}
\end{figure}

\section{Conclusions and Future Outlook}
\label{sec:discussion}

Using deployment-grade RF and readout hardware, we report full-device
NEPs of $\sim 3 \times 10^{-16}$ W\,Hz$^{-1/2}$ for our 50-channel, $R
\sim 275$ filter bank spectrometer when the device lens is illuminated
with optical loads typical of those that we will experience at the
LMT.  Most NEPs are clustered to within 20\% of the array median, so
we expect reasonably uniform on-sky sensitivity across the 255--278
GHz filter bank.  We estimate an optical efficiency through the lens
and filter bank of $\sim 10\%$.  At the tested range of loadings the
GR and photon noise components are comparable, while the contribution
from the readout system is negligible.

This demonstrated noise performance is sufficient for deployment to
the LMT.  Initially we will focus on demonstrating spectrometer
performance by detecting known CO lines in nearby galaxies and
high-$J$ CO in sources where lower-$J$ lines have been
detected. Eventually we will use SuperSpec to blindly detect [CII] in
high-$z$ galaxies, complementing the existing LMT instrumentation
suite.  In one hour we expect a detection of [CII] in a $z = 6.5$,
$L=10^{13} L_{\odot}$ galaxy at SNR $\sim 6$ (assuming a line
luminosity fraction of $1\times 10^{-3}$).

While we conservatively baseline 50-channel SuperSpec devices for our
demonstration instrument, we have also recently fabricated 100- and
300-channel devices\cite{redford18}.  If further testing in the next
few months demonstrates similar noise performance, they will be strong
candidates for deployment and will improve our spectral coverage
significantly.  Looking beyond the demonstration instrument, we
anticipate fabricating many spectrometers on single wafers within the
next year.  Such devices would be suitable for deployment in near-term
intensity mapping experiments (e.g., TIME\cite{crites14}) as a drop-in
replacement for existing grating or Fourier Transform spectrometers.

\begin{acknowledgements}
This work is supported by the National Science Foundation under Grant
No. AST-1407457.  K.~S.~Karkare is supported by the Grainger
Foundation and the Kavli Institute for Cosmological Physics at the
University of Chicago through an endowment from the Kavli Foundation
and its founder Fred Kavli.  R.~M.~J.~Janssen's research was supported
by an appointment to the NASA Postdoctoral Program at the NASA Jet
Propulsion Laboratory, administered by Universities Space Research
Association under contract with NASA.
\end{acknowledgements}

%\pagebreak

\end{document}